\documentclass[pra,twocolumn]{revtex4}
\usepackage{mathrsfs,latexsym,amsmath,psfrag,exscale,epsfig}
\newcommand{\eq}[1]{Eq.~(\ref{#1})}
\newcommand{\fig}[1]{Fig.~\ref{#1}}
\begin{document}
\title{\large\bf Surface plasma resonance in small rare gas clusters by mixing IR and
  VUV laser pulses}
\author{Christian Siedschlag}
\affiliation{FOM Institute for Atomic and Molecular Physics (AMOLF), Kruislaan
  407, 1098 SJ Amsterdam, The Netherlands}
\author{Jan M. Rost}
\affiliation{
 Max-Planck-Institute for the Physics of Complex Systems,
N\"othnitzer Str. 38, D-01187 Dresden, Germany}
\date{\today}

\begin{abstract}
The ionization dynamics of a Xenon cluster with 40 atoms is analyzed under a pump probe scenario of laser pulses
where an infrared laser pulse of 50 fs length follows with a well defined time delay a VUV pulse of the same length
and peak intensity. The mechanism of resonant energy absorption due to the coincidence of the IR laser frequency
with the frequency of collective motion of quasi free electrons in the cluster is mapped out by varying the time delay 
between the pulses.
\end{abstract}
\maketitle
In recent years, much work has been devoted to the ionization mechanisms of clusters in few-cycle, intense
laser fields (i.e. pulse lengths of the order of 100 fs and intensities
$I=10^{13} \ldots 10^{16} \mathrm{W/cm^2}$): from the case of plasmon excitation when
exposing metal clusters to relatively weak fields \cite{SuRe00} over {\em enhanced
  ionization} akin of molecular ionization for small rare gas clusters in
intense fields \cite{SiRo02} to collective excitation of a plasma resonance in
clusters of intermediate  \cite{SaRo03} to large sizes \cite{jung04}, ultimately leading to
ionic charge states of 40+ and higher \cite{ditmire}, thus potentially providing a new source
for the generation of x-rays, energetic ions or electrons and, via nuclear
fusion, even neutrons \cite{jortner}.
 A new parameter regime for  laser-cluster interaction
has been proven to become  accesible with the first experiment using VUV-FEL light of 98nm wavelength for the ionization of
rare gas clusters \cite{Waal02}, soon followed by the first proposals for an
 explanation of the unexpectedly high charge states seen in this
experiment \cite{SaGr03,SiRo04}. 

While  XUV-cluster interaction is still the
subject of an ongoing debate, there seems to be a more or less common understanding
regarding the qualitative picture of IR laser-cluster
interaction: during the rising
part of the laser pulse, a few electrons are ionized \footnote{For rare gas
  clusters,  the concepts of {\em inner} and {\em outer} ionization, the
  first being the ionization of an electron out of an atomic orbital into the
  cluster environment, the latter the process of an electron leaving the
  cluster as a whole, have proven to be very useful. If one is dealing with
  the valence electrons of a metal cluster, the inner ionization step is
  skipped, since these electrons can already move freely throughout the
  cluster. From the second shell onwards, metal clusters should not behave
  different from rare gas clusters}  leaving the cluster with a net positive charge which leads to an expansion typically
 on the same time scale as the duration of the laser pulse. Hence,
effects which depend on the internuclear distances can be resolved by varying
the pulse length \cite{kolal99} and/or applying pump-probe techniques \cite{dopal01,Zaal04}. 
The resolution of an optimum time delay $\Delta t$ and the contrast of the signal in a pump-probe experiment 
increases if  $\Delta t \gg T$, the length of each pulse. Since $\Delta t \approx t_{c}$,  the critical expansion time of the cluster
at which maximum absorption of energy from the cluster pulse is possible, long times $t_{c}$ are desirable which implies
large clusters  consisting of heavy atoms (slower Coulomb explosion). Also, the large number of quasi-free electrons temporarily trapped in the cluster, lead to a good contrast for the optimized versus non-optimized signal \cite{Zaal04}.

The critical time $t_{c}$ originates from a critical radius $R_{c} = R(t_{c})$ of the cluster, usually larger than the equilibrium 
radius $R_{0}$, where energy absorption is most efficient.
%	For
%	resonant as well as for enhanced ionization, there exists a criticial cluster
%	radius $R_{crit}$ which is usually larger than the equilibrium radius $R_0$
%	and allows for an especially efficient ionization. 
For larger clusters (resonant mechanism) this radius is determined by the
surface plasma frequency
approximately given by
\begin{align}\label{plasmon}
\Omega_{t}=\sqrt{\frac{N_{t}Z_{t}}{R_{t}^3}}= \frac{\omega_{pl}}{\sqrt{3}},
\end{align}
 where $N_{t}$ is the number of atoms/ions in the
cluster, $Z_{t}$ is their average charge, $R_{t}$ is the cluster radius and
$\omega_{pl}$ the bulk plasma frequency. 
The indices $t$ indicate a slow dependence on time (at this point we want to
emphasize that, at least as long as the cluster is neutral, the surface {\em plasma} resonance is mathematically completely
equivalent to the
surface {\em plasmon} resonance and can be derived along the same
pathway. Nevertheless, we prefer to call it a plasma resonance since there is
in principle a physical difference between a metal cluster being
excited perturbatively and a rare gas cluster turned into a nanoplasma by a
nonperturbative laser field). If $\Omega_t \approx \omega$, the laser frequency, then the cloud of
electrons which are trapped inside the cluster behaves like a (damped)
harmonic oscillator driven to resonance \cite{SaRo03}, leading to efficient energy absorption and ionization. 

The most
important prerequisite for this mechanism  is
a significant amount of trapped (quasi-free) electrons before $R_{c}$ is
reached, which can be achieved in two ways: either the laser field strength
is small enough to leave enough electrons inside the cluster
before $R=R_{c}$; this possibility is limited, however, by the fact that the
inner ionization process will eventually not start if the field strength is too
small, so that no quasi-free electrons will be created in the first place. For metal
clusters, this problem does obviously not occur; on the other hand, one has to
ionize the cluster to a certain degree in order to start the expansion
process, and the range of intensities and pulse lengths which can start the
Coulomb explosion while at the same time keeping the valence electron cloud
intact is quite small \cite{calv00}. On the other hand, the force that keeps the
electrons inside the cluster is generated by the space charge of the
ions. Hence, going to larger clusters while leaving the average ion charge
constant, will make it more and more difficult for electrons to leave the
cluster, so that the number of trapped electrons will increase with the
cluster size. This is the reason why the resonance absorption is much clearer seen with 
IR pulses for large clusters (compare \cite{dopal01} with \cite{Zaal04}).

To summarize, resonant absorption in intense IR fields occurs
(i) for metal clusters in weak fields ($I \lesssim 10^{13}
\mathrm{W/cm^2}$); the size of the clusters then only plays a role in so far as it
will change the expansion speed, which has to be accounted for by changing the
pulse lengths accordingly; or (ii) for clusters with $N \gtrsim 10^{2}$ and intensities of
$I\lesssim 10^{14} \mathrm{W/cm^2}$, where a substantial fraction of the quasi-free
electrons which are created by the laser is kept inside the cluster, so that a
collective oscillation can develop.

In the case of small rare gas clusters in strong IR fields, none of the above
scenarions applies. Rather, ionization is dominated by {\em
  charge enhanced ionization}, known already from diatomic molecules, where the
shape of the interatomic barrier leads to an optimal distance
between the cluster nuclei which results in an efficient interplay between
inner and outer ionization \cite{enhanced}: the neighbouring charges must be close
enough to start an {\em ionization avalanche} \cite{SiRo02} once the first
electrons are created; on the other hand, they must not be too close in order
to decrease the space charge which prevents the electrons from escaping the
cluster. Obviously, the whole process relies on the ionization of the first
electrons relatively early in the pulse; once the field strength drops
significantly below the field strength $F_{th}$ required to field-ionize a single
cluster atom ($F_{th}=E_b^2/4$, where $E_b$ is the first atomic ionization
energy), the avalanche will not be started and the cluster will survive the
radiation relatively undamaged. For Xe clusters, for example, $F_{th}=0.0493$
a.u., which corresponds to an intensity of $I_{th}=8.53\cdot 10^{13}
\mathrm{W/cm^2}$.

\begin{figure}
\centering
\psfrag{a}[][][1]{atom}
\psfrag{b}[][][1]{cluster}
\psfrag{xtitle}[][][1]{x [a.u.]}
\psfrag{ytitle}[][][1]{potential energy [arb. units]}
\includegraphics[scale=0.4]{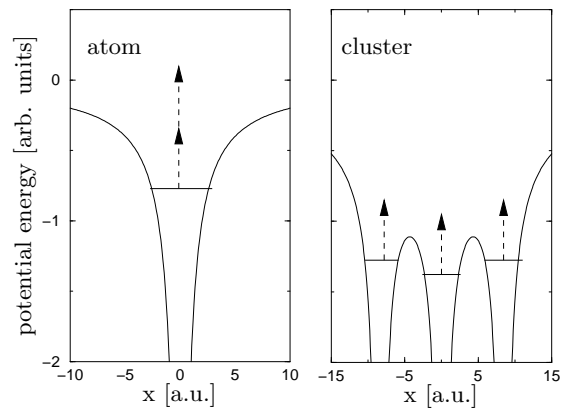}
\caption{Comparison of the inner ionization process for a single atom and a
  model cluster of three atoms. The effective barrier is lowered in a cluster
  due to the charged environment.}
\label{fig1}
\end{figure}

As has been shown in the Hamburg experiment, the threshold for an ionization avalanche is
considerably lower when using VUV instead of IR light. With intensities of the
order of $10^{12} \ \mathrm{W/cm^2}$ and a photon energy of 12.7 eV, complete
breakup of Xe clusters and unexpectedly high ionic charges have been
observed \cite{Waal02}.   
These findings can be explained \cite{SiRo04} by using standard atomic photoabsorption
rates but taking into account the effective (inner) ionization threshold which is
lowered by the surrounding charges in a cluster (see Fig \ref{fig1}). Due to this mechanism and due to the fact that the quiver amplitude is two orders of magnitude smaller than for IR radiation, a VUV pulse is much more efficient in creating quasi-free
electrons than an IR pulse of the same peak intensity. 

This opens up an elegant way to study the dynamics of collectively excited electrons and hence the resonance absorption mechanism in a  {\it small} cluster
by combining a VUV pump pulse with a time delayed IR probe pulse: The VUV pulse generates 
a large number of quasi-free electrons. At same time the cluster gets only moderately charged and
a slow expansion sets in mainly driven by the hydrodynamic pressure of the quasi-free electrons.
 Hence, one can observe with a time delayed probe pulse very cleanly the optimum condition for 
energy absorption by the quasi-free electrons as a function of cluster size starting at a size as small as $N=40$ as we will demonstrate with the following pump-probe scenario: a Xe$_{40}$ cluster is
first irradiated by a 50 fs VUV pulse ($\omega=12.7$ eV, $I=7.9\cdot10^{12} \
\mathrm{W/cm^2}$) Then, after a variable time delay $\Delta t$,
we apply a second pulse of the same length and intensity, but now with a
wavelength of 780 nm. The simulation has been done using the quasiclassical
model introduced in \cite{SiRo02}.

\begin{figure}
\centering
\psfrag{a}[][][0.9]{$\Delta t=0$ a.u.}
\psfrag{b}[][][0.9]{$\Delta t=1000$ a.u.}
\psfrag{c}[][][0.9]{$\Delta t=3500$ a.u.}
\psfrag{d}[][][0.9]{$\Delta t=8000 $ a.u.}
\psfrag{xtitle}[t][][1]{charge state}
\psfrag{ytitle}[b][][1]{counts [arb. units]}
\includegraphics[scale=0.33]{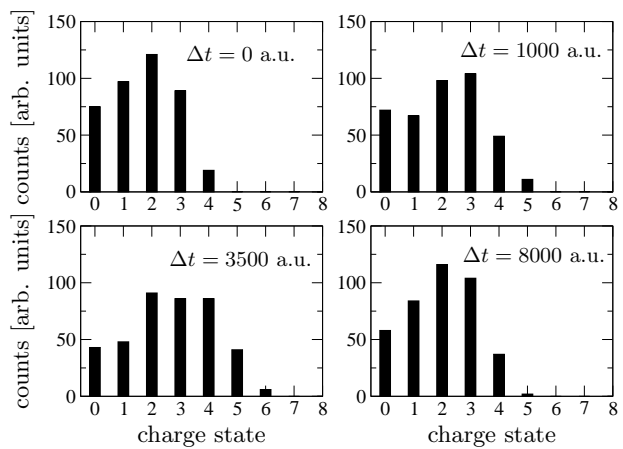}
\caption{Single ion charge spectra after the interaction of Xe$_{40}$ with a
  VUV pump pulse followed by an IR probe pulse as a function of the delay
  $\Delta t$ between the two pulses. The highest (maximum and
  average) charge states are created for $\Delta t$=3500 a.u.}
\label{fig2}
\end{figure}
The charge spectra  after the interaction of the cluster with the
two pulses is shown in Fig. \ref{fig2} for various  times delay $\Delta t$. Two
features stand out: first of all, ions with charges of $5+$ or $6+$ are produced
already with these comparatively low intensities. In fact, the
ionization efficiency is comparable to the results from our calculation in
\cite{SiRo04}, where only a single VUV pulse was applied, but with an
intensity one order of magnitude higher than in the present case (note that,
with an average charge per atom of 2.5 for $\Delta t$=3500 a.u., the space
charge of the Xe$_{40}$ cluster from the present work is approximately equal to the space charge of
the Xe$_{80}$ cluster from \cite{SiRo04}, where an average charge per atom of
only 1.5 was achieved, so that the two cases can really be compared). 
This shows that by combining VUV and IR pulses, significantly higher charge
states can be achieved than by applying an IR or a VUV pulse alone. Second, we
see that the ionization is most efficient for a time delay of $\Delta t$=3500 a.u.\\
\begin{figure}
\centering
\psfrag{xtitle}[t][][0.8]{delay [a.u.]}
\psfrag{ytitle1}[b][][0.8]{abs. energy [a.u.]}
\psfrag{ytitle2}[b][][0.8]{aver. charge/atom}
\psfrag{a}[][][1]{a)}
\psfrag{b}[][][1]{b)}
\includegraphics[scale=0.33]{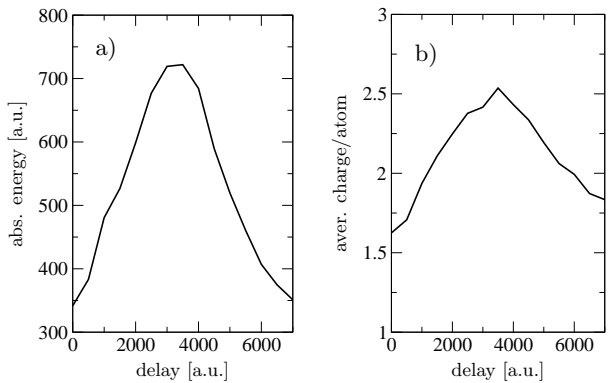}
\caption{Absorbed energy [a)] and average charge per atom [b)] as a function
  of time delay between the VUV and the IR pulse.}
\label{fig3}
\end{figure}
The absorbed energy and the average charge state per atom after the cluster
has disintegrated are shown in \fig{fig3} as a function of the  time delay $\Delta t$. The curve for the energy
absorption as well as the one for the charge shows a maximum 
$\Delta t \approx$ 3500 a.u. between the pump VUV and the probe IR pulse. 
If our physical picture is correct, this maximum should be due to the existence of   a
collective resonance with frequency $\Omega_{t}$ of the quasi-free electrons that were created by the VUV
pulse. This resonance is most efficiently excited by the IR pulse when the laser frequency $\omega=\Omega_{t}$.
 We will now proceed to
give evidence for this hypothesis.\\
In principle there are two ways to check numerically whether the electron
cloud is at resonance with the laser field: first, treating the electrons and
ions of the cluster as a homogenic positively and negatively charged sphere,
respectively, one ends up with the  resonance condition $\Omega_{t}=\omega$.
However, the definition \eq{plasmon} of $\Omega_{t}$ is by no means unique,
since one has to define a cluster ``volume'' which itself is time dependent (through the increase of the cluster radius) and so are the charge of the ions and the number of electrons. The resonance condition  can be determined 
more reliably by calculating the phase difference between the oscillation of the electronic
center of mass (ECM) and the driving laser field $F(t) = F_{t}\cos\omega t$  \cite{SaRo03}: if one assumes a collective
oscillation with a damping constant $\gamma$, the time-dependent dipole
amplitude for the ECM reads
\begin{align}
X(t)=A_t \cos(\omega-\phi_t)
\end{align}
with
\begin{align}\label{amplitude}
A_t&=F_{t}/\sqrt{(\Omega_t^2-\omega^2)^2+(2\Gamma_t\omega)^2}\\
\phi_t&=\arctan(2\Gamma_t\omega/(\Omega_t^2-\omega^2)).
\end{align}
For $\phi_t=\pi/2$ the system is at resonance and the laser cycle averaged energy absorption
\begin{align}
\langle dE/dt\rangle = \frac 1T\int_{0}^{T}\frac{dX_{t}}{dt}F(t)\,dt\propto \sin\phi_{t}
\end{align}
is at its maximum. Note, however, that the amplitude $X_{t}$ does not necessarily 
increase considerably at resonance due to strong damping (see \eq{amplitude}).
Hence, we take the condition 
$\phi_t=\pi/2$ as the {\em definition} for the plasmon resonance and calculate $\phi_{t}$ from
the phase lag between the driving field $F(t)$ and the dipole
oscillation $dX_{t}/dt$ by extracting the maximum of the time correlation $c(\delta t)$ between
the two signals, where
\begin{align}
c(\delta t)=\int_{t_1}^{t_2} F(t) X(t+\delta t) \, dt
\label{corr_eq}
\end{align}
(note that $\delta t \neq \Delta t$!). 
We chose the limits of integration in \eq{corr_eq} to be 10 IR cycles
before and after the maximum of the probe laser; the radial evolution $R_{t}$ of the
cluster is sufficiently slow so that the phase lag only changes by a small
amount during that time. The outcome of this calculation is shown in
Fig. \ref{fig4}. Indeed the phase lag is equal to $\pi/2$ for $\Delta t\approx
3800 a.u.$. This proves that the maxima in Fig. \ref{fig3} are due to a
resonance of the collective electron oscillation with the driving laser
field. 

Hence, the VUV pump combined with a IR probe pulse can map out the internal collective cluster dynamics
very clearly and may be the only possibility to resolve this dynamics for small  clusters. The reason is simply that the VUV pulse produces a large number of quasi-free electrons which can participate in collective electron motion. On the other hand the VUV pulse itself does not couple to this collective  motion. Hence, only the probe pulse probes literally the collective electron dyamics. Using two IR pulses for pump and probe make this distinction difficult:
The pump pulse must be very short in order not to ``probe'' itself the collective electron dynamics. On the other hand, if it  is very short it will not produce a significant number of quasi-free electron which could move collectively since they are produced most efficiently through resonant coupling.
In the next phase of the VUV-FEL at Hamburg, a pump-probe facility as ``used''
in this theoretical investigation will be available for
experiments. Furthermore a similar experiment, but much simpler than the one
at DESY as far as the experimental set is concerned, is planned in Saclay
\cite{hubertus}; it is planned to use an 800 nm femtosecond laser,
together with its 9th harmonic (delivering photons with $E\approx$ 14 eV) as
the IR and VUV pulse, respectively; the 
intensities will be a bit lower than the ones considered here, but this should
only result in a quantitative difference. The present
paper shows that indeed, interesting and unique experiments can be done with
such a time-delayed combination of pulses.\\
This work is part of the research program of FOM (Fundamental Research on
Matter), which is subsidized by the NWO (Netherlands Organization for the
Advancement of Research).
    
\begin{figure}
\centering
\psfrag{xtitle}[t][][1]{delay [a.u.]}
\psfrag{ytitle}[b][][1]{phase lag}
\psfrag{p1}[][][1]{$\frac{\pi}{2}$}
\psfrag{p2}[][][1]{$\pi$}
\includegraphics[scale=0.33]{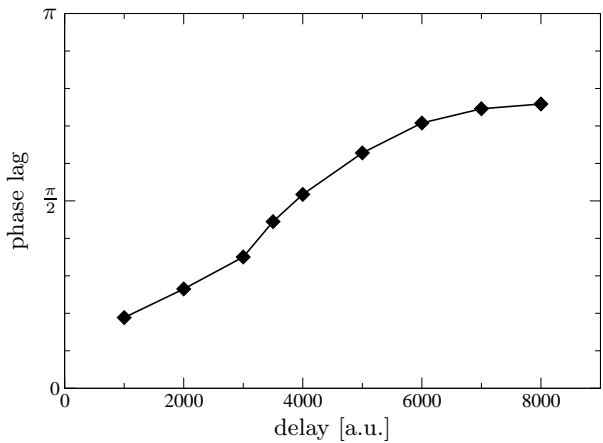}
\caption{Phase lag between the center-of-mass oscillation of the electron
  cloud and the driving laser field as a function of time delay between the
  VUV and the IR pulse.}
\label{fig4}
\end{figure}

\bibliographystyle{unsort}

\begin{thebibliography}{99}
\bibitem{SuRe00}  E.~Suraud and P.~G. Reinhard, Phys. Rev. Lett. {\bf 85}, 2296 (2000).
\bibitem{enhanced} T. Seideman, M.Y. Ivanov and P.B. Corkum, Phys. Rev. Lett. {\bf 75}, 2819 (1995).
\bibitem{SiRo02} Ch. Siedschlag and J.M. Rost, Phys.\ Rev.\ Lett. {\bf 89}, 173401
(2002); Phys.~Rev.~A {\bf 67}, 013404 (2003).
\bibitem{SaRo03} U. Saalmann and J.M. Rost, Phys.\ Rev.\ Lett. {\bf 91}, 223401 (2003).
\bibitem{jung04} C. Jungreuthmayer, M. Geissler, J. Zanghellini and T. Brabec,
  Phys.\ Rev.\ Lett. {\bf 92}, 133401 (2004).
\bibitem{ditmire} T. Ditmire, T. Donnelly, A.M. Rubenchik, R.W. Falcone and
  M.D. Perry, Phys.\ Rev.\ A {\bf 53}, 3379 (1996).
\bibitem{jortner} I. Last and J. Jortner, J.\ Phys.\ Chem.\ A {\bf 106},
  10872 (2002).
\bibitem{Waal02} H. Wabnitz {\em et. al.}, Nature {\bf 420}, 482 (2002).
\bibitem{SaGr03} R. Santra and C. H. Greene, Phys.\ Rev.\ Lett. {\bf 91},
  233401 (2003).
\bibitem{SiRo04} Ch. Siedschlag and J.M. Rost, Phys.\ Rev.\ Lett. {\bf 93},
  043402 (2004).
\bibitem{kolal99} L.~K{\"o}ller, M.~Schumacher, J.~K{\"o}hn, S.~Teuber,  J.~Tigges\-b{\"a}umker, and K.~H. Meiwes-Broer, 
  Phys. Rev. Lett. {\bf 82}, 3783 (1999).
\bibitem{dopal01} T. D\"oppner, Th. Diederich, J. Tiggesb\"aumker, and K.ÐH. MeiwesÐBroer, Eur. Phys. J. D {\bf16}, 13 (2001).
\bibitem{Zaal04} S. Zamith, T. Martchenko, Y. Ni, S. A. Aseyev, H. G. Muller, and M. J. J. Vrakking, Phys.\  Rev. {\bf A 70}, 011201(R) (2004).
\bibitem{calv00} F. Calvayrac, P.-G. Reinhard, E. Suraud and C.A. Ullrich, Phys.\ Rep. {\bf 337}, 493 (2000)
\bibitem{hubertus} H. Wabnitz, private communication.
\end{thebibliography}

\end{document}